\documentclass[twocolumn]{aastex701}
\usepackage{amsmath}
\usepackage{xspace}

\makeatletter
\AtBeginDocument{\def\Hy@footnote@currentHref{titlefootnote}}
\makeatother

\newcommand{\sacra}{\texttt{SACRA}\xspace}
\newcommand{\sacrampi}{\texttt{SACRA-MPI}\xspace}
\newcommand{\sacrak}{\texttt{SACRA-K}\xspace}
\newcommand{\fuka}{\texttt{FUKA}\xspace}
\newcommand{\kokkos}{\texttt{Kokkos}\xspace}
\newcommand{\nanasi}{\texttt{NANASI}\xspace}

\begin{document}
 
\title{\texttt{SACRA-K}\xspace: A Performance-Portable Numerical Relativity Code with \texttt{Kokkos}\xspace}

\author[orcid=0000-0001-9034-0866,sname='Han']{Ming-Zhe Han}
\affiliation{Max-Planck-Institut f\"ur Gravitationsphysik (Albert-Einstein-Institut), Am M\"uhlenberg 1, D-14476 Potsdam-Golm, Germany}
\affiliation{Key Laboratory of Dark Matter and Space Astronomy, Purple Mountain Observatory, Chinese Academy of Sciences, Nanjing, 210033, People's Republic of China}
\email[show]{ming-zhe.han@aei.mpg.de}  

\author{Kenta Kiuchi}
\affiliation{Max-Planck-Institut f\"ur Gravitationsphysik (Albert-Einstein-Institut), Am M\"uhlenberg 1, D-14476 Potsdam-Golm, Germany}
\affiliation{Center for Gravitational Physics and Quantum Information, Yukawa Institute for Theoretical Physics, Kyoto University, Kyoto, 606-8502, Japan}
\email{kenta.kiuchi@aei.mpg.de}

\author{Masaru Shibata}
\affiliation{Max-Planck-Institut f\"ur Gravitationsphysik (Albert-Einstein-Institut), Am M\"uhlenberg 1, D-14476 Potsdam-Golm, Germany}
\affiliation{Center for Gravitational Physics and Quantum Information, Yukawa Institute for Theoretical Physics, Kyoto University, Kyoto, 606-8502, Japan}
\email{mshibata@aei.mpg.de}

\date{\today}

\begin{abstract}
We present \texttt{SACRA-K}\xspace (SimulAtor for Compact objects in Relativistic Astrophysics with \texttt{Kokkos}\xspace), a performance-portable numerical relativity code ported from the Fortran code \texttt{SACRA-MPI}\xspace to C++ with the \texttt{Kokkos}\xspace library.
\texttt{SACRA-K}\xspace retains the physics and numerical methods of \texttt{SACRA-MPI}\xspace, namely a Baumgarte-Shapiro-Shibata-Nakamura (BSSN) spacetime evolution with Z4c constraint propagation and a box-in-box Berger-Oliger adaptive mesh refinement, together with the high resolution shock capturing scheme for the general relativistic hydrodynamics from \texttt{NANASI}\xspace, while gaining cross vendor portability.
We validate the port against \texttt{SACRA-MPI}\xspace across binary black hole, black hole neutron star, and binary neutron star configurations.
Across these tests, the waveform discrepancies are well below both the scatter among independent numerical relativity codes and the resolution dependent variation within a given code, and remain at or below the distinguishability threshold of current gravitational-wave detectors; the $\pi$ symmetry is preserved exactly at the bitwise level; and the gravitational wave phase of the binary neutron star merger exhibits second order convergence.
In the smallest test configuration, \texttt{SACRA-K}\xspace runs about an order of magnitude faster on the NVIDIA A100 GPU cluster or the AMD MI300A APU cluster than the Fortran \texttt{SACRA-MPI}\xspace on the CPU cluster, and we measure its scaling up to 256 accelerator devices.
\end{abstract}

\section{Introduction}\label{sec:intro}

The direct detection of gravitational waves (GWs) has opened a new observational window on the Universe, beginning with GW150914 from a binary black hole (BBH) merger \citep{LIGOScientific:2016aoc} and, soon afterwards, GW170817 from a binary neutron star (BNS) merger \citep{LIGOScientific:2017vwq}.
The latter opened the era of multimessenger astronomy.
GW170817 was observed together with a short gamma ray burst and the kilonova AT2017gfo \citep{LIGOScientific:2017ync}.
Its emission was likely to be powered by the radioactive decay of $r$ process nuclei \citep{Kasen:2017sxr}, and it provided evidence for the hypothesis long anticipated~\citep{Li:1998bw,Metzger:2010sy}.
These observations translate directly into physics.
The tidal deformability measured from GW170817 constrains the neutron star (NS) radius and the equation of state (EOS) of dense matter \citep{LIGOScientific:2018cki}.
Kilonovae, in turn, link the GW and electromagnetic signals to the origin of the heavy elements \citep{Shibata:2019wef, Metzger:2019zeh}.

Extracting this physics from the observations requires an accurate model of the binary dynamics.
During the inspiral, the dynamics can be described by analytic approximations, namely the post Newtonian (PN) \citep{Blanchet:2013haa} and effective one body (EOB) \citep{Buonanno:1998gg} methods.
At merger, however, the binary enters a strong field, highly nonlinear regime.
Only numerical relativity (NR) can follow it, through a full solution of the $3+1$ Einstein equations coupled to general relativistic hydrodynamics \citep{2016nure.book.....S,Dietrich:2020eud}.
Moreover, the PN and EOB models themselves must be calibrated against NR.
The tidal waveform models used to interpret BNS signals, both the phenomenological NRTidal models \citep{Dietrich:2017aum} and the tidal EOB models \citep{Bernuzzi:2014owa,Kawaguchi:2018gvj}, rely on NR for calibration. 
The parameter estimation of GW170817 itself made direct use of such models \citep{LIGOScientific:2018hze}.
NR also reveals dynamics that the observations cannot resolve on their own.
These range from the mass ejection and $r$ process nucleosynthesis in the merger remnant \citep{Shibata:2019wef,Fujibayashi:2022ftg,Kiuchi:2022nin} to subtle effects such as the $f$ mode tidal resonance, whose numerical study exposes the limitations of analytic models \citep{Kuan:2024jnw}.

This power, however, comes at a steep computational cost.
A single high resolution BNS simulation that resolves the waveform to sub radian accuracy already costs hundreds of thousands of core hours \citep{Kiuchi:2017pte}.
Mapping the physics across the EOS, mass, mass ratio, and spin requires systematic surveys of many such runs \citep{Kiuchi:2019kzt}.
The resolution itself decides which physics can be captured.
Resolving the magnetorotational instability and the magnetic field amplification in the remnant demands grid spacings of order ten meters or less, at a correspondingly enormous cost \citep{Kiuchi:2023obe}.
The next generation of detectors makes this problem even more acute.
Reaching the accuracy required by the Einstein Telescope \citep{ET:2019dnz} and Cosmic Explorer \citep{Evans:2021gyd} will demand a several fold increase in resolution, and simulations of order a million core hours each \citep{Habib:2025bkb}.
The community has therefore argued that exascale class computing is needed to meet these demands \citep{Foucart:2022iwu}.

Meeting this demand, in turn, requires a fundamental change in computing hardware.
As the growth of single thread performance has slowed, the leading high performance computing systems have moved to heterogeneous architectures built around graphics and accelerated processing units (GPUs and APUs).
The vendor specific programming models for these accelerators have, however, severely fragmented the software landscape.
Performance portability has therefore become essential: the ability to run a single source code efficiently across central processing units (CPUs), GPUs, and APUs from different vendors.
Abstraction layers such as \kokkos \citep{CARTEREDWARDS20143202,dbabca8444814d58bb3dc26c5f92ebe6} provide it, together with related frameworks such as AMReX \citep{2020arXiv200912009Z} and Parthenon \citep{2022arXiv220212309G}.
The viability of \kokkos against other programming models has been carefully assessed \citep{2024arXiv240208950D}, and portability itself has been given a precise definition and metric \citep{2016arXiv161107409P}.
A direct precedent for our own approach is K-Athena \citep{2021ITPDS..32...85G}.
It ported the established Athena++ code to \kokkos and demonstrated portable performance on both CPUs and GPUs.
The NR community has responded with a wave of GPU accelerated codes.
AthenaK, built on \kokkos \citep{2026ApJS..283...27S}, provides a vacuum spacetime module that reaches close to two orders of magnitude speedup over a single CPU core and scales across tens of thousands of GPUs \citep{Zhu:2024utz}, together with a performance-portable BNS magnetohydrodynamics (MHD) module \citep{Fields:2024pob}.
Other efforts build on AMReX/CarpetX, including GRaM-X \citep{Shankar:2022ful}, AsterX \citep{Kalinani:2024rbk}, which runs about an order of magnitude faster on a GPU node than on a CPU node, and the GPU port of the subcycling time integration in CarpetX \citep{Ji:2025rpd}.
Still other routes adopt wavelet adaptive meshes on GPUs, as in Dendro-GR \citep{Fernando:2022php}, or \kokkos based general relativistic MHD, as in KHARMA \citep{Prather:2024hsu} and Phoebus \citep{Barker:2024afq}.

Against this backdrop, \sacra \citep{Yamamoto:2008js} likewise calls for a move toward GPU/APU acceleration and performance portability.
This code introduced a box-in-box adaptive mesh refinement (AMR) formulation and a hybrid MPI+OpenMP parallelization.
It has since matured into a BNS and black hole neutron star (BHNS) simulation program that produces sub radian accuracy waveforms for GW data analysis \citep{Kiuchi:2017pte,Kiuchi:2019kzt}.
The code, however, is written in Fortran for CPUs.
It does not exploit the GPU and APU accelerators that now dominate the largest machines.
This gap motivates the present work.

In this work, we present \sacrak, a C++ port of \sacrampi built on the \kokkos performance portability library \citep{CARTEREDWARDS20143202,dbabca8444814d58bb3dc26c5f92ebe6}.
A single source code can then run efficiently across CPUs, GPUs, and APUs.
By design, \sacrak retains the physics and numerical methods of the \sacrampi lineage while gaining cross vendor portability.
It shares the \kokkos based approach of AthenaK \citep{Zhu:2024utz,Fields:2024pob,2026ApJS..283...27S}, but it ports an established and well tested code rather than building one from the ground up.
We validate the port against \sacrampi, finding (i) close agreement in the waveforms, (ii) exact bitwise preservation of the $\pi$ symmetry, and (iii) second order convergence.
We then measure its performance.
The smallest configuration with 4 accelerators runs about an order of magnitude faster than the Fortran code on 1 node in a typical CPU cluster with 72 cores, and the code scales well when the total problem grows with the device count.

Throughout this paper, we adopt geometric units with $c = G = 1$, where $c$ and $G$ denote the speed of light in vacuum and gravitational constant, respectively.
Greek indices $\mu, \nu, \cdots$ run over the spacetime components $\{0, 1, 2, 3\}$, and Latin indices $i, j, \cdots$ over the spatial components $\{1, 2, 3\}$.

The remainder of this paper is organized as follows.
In Section~\ref{sec:methodology} we describe the formulation, numerical methods, and implementation; in Section~\ref{sec:validation} we present the validation tests; in Section~\ref{sec:performance} we report the performance; and finally, we summarize in Section~\ref{sec:summary}.

\section{Methodology}\label{sec:methodology}

In this section, we describe the theoretical and numerical methods used in \sacrak.
More details can be found in \citet{Yamamoto:2008js,2016nure.book.....S,Kiuchi:2022ubj}.

\subsection{Evolution equations}\label{sec:method_evolution}

\subsubsection{Spacetime evolution}\label{sec:method_spacetime}

\sacrak evolves the spacetime geometry with the same Baumgarte-Shapiro-Shibata-Nakamura (BSSN) \citep{Shibata:1995we,Baumgarte:1998te} puncture \citep{Baker:2005vv,Campanelli:2005dd} formulation as \sacrampi \citep{Yamamoto:2008js}.
This recasts the $3+1$ spatial metric $\gamma_{ij}$ and extrinsic curvature $K_{ij}$ into the conformal set of evolved variables
\begin{align}
W &= (\det\gamma_{ij})^{-1/6}, \\
\tilde\gamma_{ij} &= W^2\gamma_{ij}, \\
K &= \gamma^{ij}K_{ij}, \\
\tilde A_{ij} &= W^2\Big(K_{ij} - \tfrac{1}{3}\gamma_{ij}K\Big), \\
\tilde\Gamma^i &= -\partial_j\tilde\gamma^{ij},
\end{align}
namely the conformal factor $W$, the conformal spatial metric $\tilde\gamma_{ij}$, the trace of the extrinsic curvature $K$, its conformal traceless part $\tilde A_{ij}$, and the conformal connection function $\tilde\Gamma^i$.
\sacrak adopts Cartesian coordinates.
In these coordinates the conformal metric has unit determinant by construction, $\det\tilde\gamma_{ij} = 1$, and $\tilde A_{ij}$ is traceless with respect to it.
To control the growth of the constraint violations during the evolution, we locally apply the Z4c prescription to propagate them away \citep{Hilditch:2012fp}.
This introduces the Z4 scalar $\Theta$ as an additional evolved field, with damping parameters $\kappa_1$ and $\kappa_2$.
The coordinates are fixed by the moving puncture gauge.
The lapse $\alpha$ obeys the 1+log slicing condition,
\begin{equation}
(\partial_t - \beta^j\partial_j)\alpha = -2\alpha K ,
\end{equation}
and the shift $\beta^i$ follows a hyperbolic Gamma driver condition,
\begin{align}
(\partial_t - \beta^j\partial_j)\beta^i &= 0.75B^i ,\\
(\partial_t - \beta^j\partial_j)B^i &= (\partial_t - \beta^j\partial_j)\tilde\Gamma^i - \eta B^i ,
\end{align}
where $B^i$ is an auxiliary field and $\eta$ a damping parameter.
The full set of evolved geometric variables is therefore $W$, $\tilde\gamma_{ij}$, $K$, $\tilde A_{ij}$, $\tilde\Gamma^i$, and $\Theta$, together with the gauge variables $\alpha$, $\beta^i$, and $B^i$.

\subsubsection{Matter evolution}\label{sec:method_matter}

\sacrak evolves the matter as a perfect fluid without magnetic fields, see \nanasi \citep{Kiuchi:2022ubj} for details.
The stress energy tensor is written as
\begin{equation}
T_{\mu\nu} = \rho h\, u_\mu u_\nu + P g_{\mu\nu},
\end{equation}
where $g_{\mu\nu}$ is the four-dimensional metric, $\rho$ is the rest mass density, $u^\mu$ the four velocity, $P$ the pressure, $h$ the specific enthalpy, defined by
\begin{equation}
h = 1 + \varepsilon + \frac{P}{\rho},
\end{equation}
and $\varepsilon$ the specific internal energy.
The evolved matter variables are the conserved rest mass density, momentum density, and energy density,
\begin{equation}
D = \rho w, \qquad J_i = \rho h w\, u_i, \qquad \rho_\mathrm{H} = \rho h w^2 - P,
\end{equation}
where $w = \alpha u^t$ is the Lorentz factor measured by the Eulerian observer.
The system is closed by a hybrid EOS,
\begin{equation}
P = P_{\rm cold}(\rho) + (\Gamma_{\rm th} - 1)\,\rho\,[\varepsilon - \varepsilon_{\rm cold}(\rho)] .
\end{equation}
Its cold part $\{P_{\rm cold}, \varepsilon_{\rm cold}\}$ is supplied either in tabulated form or as a piecewise polytrope.
Its thermal part is governed by the constant thermal index $\Gamma_{\rm th}$.

\subsection{Numerical methods}\label{sec:method_numerics}

In \sacrak, we advance the spacetime and matter variables together by the method of lines.
The spatial and temporal discretizations are therefore treated separately.
The time integration uses the classical fourth order Runge-Kutta scheme.
The spatial discretization differs between the two sectors, namely finite differencing for the spacetime variables and a high resolution shock capturing finite volume scheme for the matter.

The geometric variables are discretized on the Cartesian grid by finite differencing.
We use fourth order centered differences for spatial derivatives.
For the advection terms along the shift, we instead use a fourth order upwind stencil, with a bias that follows the sign of the shift.
To suppress high frequency grid noise, we add sixth order Kreiss-Oliger (KO) dissipation \citep{kreiss1973methods} to the evolved metric variables.
The dissipation coefficient is set to $\sigma_{\rm KO}=0.5$.
The time step of each refinement level is set by a Courant factor of $0.5$.

The finite volume scheme reconstructs the primitive variables at the cell interfaces with a third order piecewise parabolic method (PPM), limited by a steep minmod limiter with parameter $b = 3$ \citep{Colella:1982ee}.
The interface fluxes are computed with a Harten-Lax-van Leer contact (HLLC) approximate Riemann solver \citep{Mignone:2005ft}.
It reduces to the Harten-Lax-van Leer-Einfeldt (HLLE) \citep{DelZanna:2002rv} solver wherever the intermediate state would be unphysical, namely at a negative intermediate pressure or a superluminal contact speed.
The Riemann problem is solved in a local Minkowski frame obtained by a tetrad transformation.
The special relativistic solver can then be applied directly at each interface.
After each update the primitive variables $(\rho, \varepsilon, u_i)$ are recovered from the conserved ones by a Newton-Raphson iteration on the Lorentz factor $w$ (see section 4.4.1 in \citet{2016nure.book.....S} for more details.).
A cell is reset to a tenuous atmosphere at rest if it falls below the density floor, or if the recovered primitive state is unphysical.
Inside a central region that extends to eight times the size of the finest domain, the floor is set to $10^{-12}$ of the maximum rest mass density; outside, it decreases as $r^{-3}$, where $r$ is the coordinate radius.

\subsubsection{Grid structure and AMR}\label{sec:method_amr}

\sacrak uses the same box-in-box Berger-Oliger adaptive mesh refinement \citep{berger1989local} hierarchy as \sacrampi \citep{Yamamoto:2008js}, with time subcycling.
This concentrates the resolution around the compact objects.
Each finer level is advanced with a halved time step and therefore takes twice as many substeps per coarse step.

\begin{figure*}[t]
    \centering
    \includegraphics[width=0.95\textwidth]{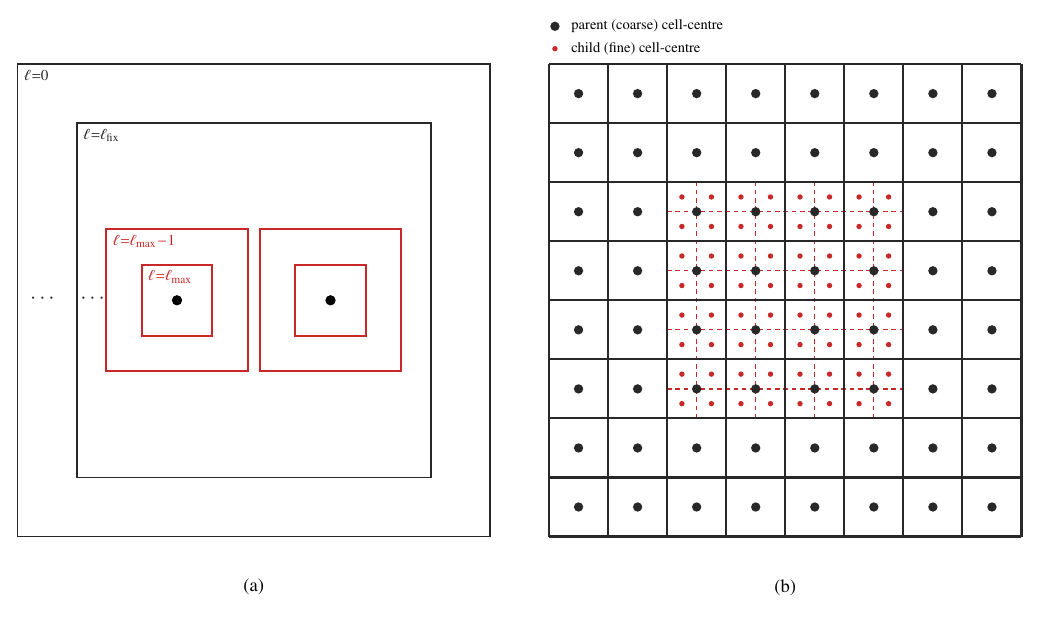}
    \caption{Schematic figures of the AMR scheme in \sacrak. 
    \emph{Panel (a):} hierarchical computational domain structure. 
    For $0 \le \ell \le \ell_{\rm fix}$ each level consists of a single cubic patch centered on the coordinate origin (\emph{fixed levels}, black). 
    For $\ell_{\rm fix} < \ell \le \ell_{\max}$ each level consists of two patches that follow the two compact objects (\emph{moving levels}, red). 
    Successive levels are related by a $2{:}1$ refinement ratio, so the grid spacing halves while the number of cells per patch is held fixed.
    \emph{Panel (b):} cell centered geometry across a refinement boundary.
    Parent (coarse) cell centers (large black dots) and child (fine) cell centers (small red dots) are mutually offset by $\Delta x/4$
    in each direction.}
    \label{fig:amr_overview}
\end{figure*}

Figure~\ref{fig:amr_overview} shows the AMR structure of \sacrak.
The hierarchy consists of (i) fixed levels, each a single cubic patch centered on the coordinate origin, and (ii) moving levels, each a pair of patches that track the two compact objects.
Successive levels are related by a $2{:}1$ refinement ratio on a cell centered staggered grid.
The grid spacing therefore halves from each coarse (parent) level to the next finer one (child), while the number of cells per patch is held fixed.
Moreover, each subdomain is padded by a buffer of six cells on every side.

Data move between levels through three operations: prolongation, restriction, and exchange.
Prolongation fills the refinement boundary zones of a child level, together with the regions newly covered when a moving patch shifts.
It interpolates from the parent level with a sixth order Lagrange interpolation.
In time, the refinement boundary data needed at the substeps are interpolated between the parent time levels, at first order in the very first time step of a run and at second order thereafter.
The conserved rest mass density is instead prolonged with a conservative fifth order weighted essentially non oscillatory (WENO5) interpolation \citep{1996JCoPh.126..202J}.
Wherever the high order interpolation gives an unphysical value, the conserved rest mass density falls back to the parent cell value.
The specific energy and enthalpy then fall back to a trilinear interpolation.
Restriction performs the reverse transfer.
It overwrites each coarse cell covered by a finer level with a fourth order Lagrange interpolation of the fine data.
The conserved variables are instead restricted by a volume average over the overlapping fine cells, which maintains conservation.
The exchange makes the different levels at the same depth consistent with one another in their overlap regions.
The result depends on which region does the overlapping.
If an active region overlaps another active region, the field data are averaged.
If an active region overlaps a refinement boundary region, it overwrites that region.
If a refinement boundary region does the overlapping, nothing happens.

The moving levels are regridded as the compact objects move.
Specifically, the location of a black hole is obtained by integrating the puncture motion, $d x^i/dt = -\beta^i$.
That of a neutron star is obtained by following its rest mass density maximum.
Each moving patch is then shifted in steps of two cells.
This keeps the surfaces of parent and child cells overlapped.
A region newly covered by a shifted patch is filled by the prolongation from the parent level.

At every boundary between a coarse and a fine level we replace the coarse flux by the sum of the fine fluxes across the matching interface.
This reflux operation, which was not installed in the original version of \sacra \citep{Yamamoto:2008js}, enforces the conservation of rest mass, momentum, and energy across the boundary.

\subsubsection{Boundary conditions}\label{sec:method_boundary}

At the outer boundary of the coarsest refinement level we impose an outgoing wave condition on the metric variables, written in the Sommerfeld form
\begin{equation}
\partial_t(rQ) + \partial_r(rQ) = 0,
\end{equation}
where $Q$ is each evolved metric variable, measured relative to its flat space value.
This lets $rQ$ propagate outward at the speed of light.
The variable $Q$ therefore falls off as $1/r$ in the wave zone.

For the configurations evolved in this work, we impose reflection symmetry with respect to the equatorial plane $z = 0$.
We thus evolve only the upper half space $z \ge 0$.

\subsection{Diagnostics}\label{sec:method_diagnostics}

We extract the gravitational waveforms from the complex Weyl scalar $\Psi_4$,
\begin{equation}
\Psi_4(t_{\rm ret}, r_0, \theta, \phi) = \sum_{l,m} \Psi_4^{l,m}(t_{\rm ret}, r_0)\, 
{}_{-2}Y_{lm}(\theta,\phi),
\end{equation}
where ${}_{-2}Y_{lm}$ is the spin weight $-2$ spherical harmonic and $t_{\rm ret}$ the retarded time.
We extract $\Psi_4^{l,m}$ on a sphere of fixed radius $r_0$ and extrapolate it to infinity by Nakano's method \citep{Nakano:2015rda},
\begin{equation}
\begin{split}
D_0\Psi_4^{l,m,\infty}(t_{\rm ret}) \equiv C(r_0)\bigg[ & D_0\Psi_4^{l,m}(t_{\rm ret}, r_0) - \frac{(l-1)(l+2)}{2} \\
& \times \int^{t_{\rm ret}} \Psi_4^{l,m}(t', r_0)\,dt' \bigg],
\end{split}
\end{equation}
with $D_0 \approx r_0[1 + m_0/(2r_0)]^2$ and $C(r_0) = 1 - 2m_0/D_0$ \citep{Kiuchi:2017pte,Kiuchi:2019kzt}, where $m_0 \equiv m_1 + m_2$ is the total mass, and the retarded time
\begin{equation}
t_{\rm ret} \equiv t - \left[D_0 + 2m_0\ln\!\left(\frac{D_0}{2m_0} - 1\right)\right].
\end{equation}
The GW strain follows from the double time integral of $\Psi_4^{l,m,\infty}$,
\begin{equation}
\begin{split}
h^{l,m,\infty}(t_{\rm ret}) &= h_+^{l,m,\infty}(t_{\rm ret}) - i\,h_\times^{l,m,\infty}(t_{\rm ret}) \\
&= -\int^{t_{\rm ret}} dt' \int^{t'} \Psi_4^{l,m,\infty}(t'')\,dt'',
\end{split}
\end{equation}
which we evaluate by the fixed frequency integration method \citep{Reisswig:2010di},
\begin{equation}
\begin{split}
h^{l,m,\infty}(t_{\rm ret}) &= \int df'\,\frac{\tilde\Psi_4^{l,m,\infty}(f')}{(2\pi\max[|f'|, f_{\rm cut}])^2} \\
&\quad\times e^{2\pi i f' t_{\rm ret}},
\end{split}
\end{equation}
where $\tilde\Psi_4^{l,m,\infty}(f)$ is the Fourier transform of $\Psi_4^{l,m,\infty}(t)$ and the cutoff frequency is $f_{\rm cut} = 0.8\,m\Omega_0/(2\pi)$, with $\Omega_0$ the initial orbital angular velocity.
Each mode is finally written in amplitude and phase form,
\begin{equation}
h^{l,m,\infty}(t_{\rm ret}) = A^{l,m,\infty}(t_{\rm ret})\,e^{-i\phi_{\rm GW}^{l,m}(t_{\rm ret})},
\end{equation}
of which the dominant $l = m = 2$ mode is used in the validation tests of Section~\ref{sec:validation}.

We monitor the Hamiltonian and momentum constraints at every step,
\begin{align}
\mathcal{H} &= R + K^2 - K_{ij}K^{ij} - 16\pi\rho_{\rm H}, \\
\mathcal{M}^i &= D_j\!\left(K^{ij} - \gamma^{ij}K\right) - 8\pi \gamma^{ij}J_j,
\end{align}
where $R$ and $D_j$ are the Ricci scalar and covariant derivative of $\gamma_{ij}$, respectively.
For each constraint $\mathcal C \in \{\mathcal H,\,\mathcal M^x,\,\mathcal M^y,\,\mathcal M^z\}$ we output its $L^2$ norm normalized by the total volume,
\begin{equation}
\|\mathcal C\|_2 = \left(\frac{\sum_{\rm cells}\mathcal C^2\,\Delta V}{\sum_{\rm cells}\Delta V}\right)^{1/2},
\end{equation}
where $\Delta V$ is the cell volume.

We also track several conserved and global quantities.
The Arnowitt-Deser-Misner (ADM) mass and the ADM-like angular momentum are evaluated as surface integrals on the coordinate spheres in the wave zone,
\begin{align}
M_{\rm ADM}(r) &= \frac{1}{16\pi}\oint_{S_r} \gamma^{ik}\gamma^{jl}\bigl(\partial_j\gamma_{kl} - \partial_k\gamma_{jl}\bigr)\,dS_i, \\
J^{\rm ADM}_z(r) &= \frac{1}{8\pi}\oint_{S_r}\bigl(x\,K_y{}^j - y\,K_x{}^j\bigr)\,dS_j,
\end{align}
where $dS_i$ is the proper surface element of the sphere.
The total baryon mass is tracked as a measure of mass conservation,
\begin{equation}
M_{\rm b} = \int \sqrt{\gamma}D\,d^3x.
\end{equation}

\subsection{Implementation of \kokkos}\label{sec:method_implementation}

\sacrak is a C++ port of \sacrampi built for performance portability, namely a single source code that runs efficiently on CPUs, GPUs, and APUs.
Accelerators from different vendors are otherwise programmed through incompatible, vendor specific models such as CUDA, HIP, and SYCL.
Maintaining a separate version of the code for each architecture would be time consuming and error prone.
We therefore build \sacrak on the \kokkos library \citep{CARTEREDWARDS20143202,dbabca8444814d58bb3dc26c5f92ebe6}, an abstraction layer that maps the same source onto the native backend of each architecture at compile time.
The backends are CUDA for NVIDIA GPUs, HIP for AMD GPUs and APUs, and OpenMP for CPUs.
We compile with \texttt{-fno-fast-math} on every platform to avoid aggressive floating point optimizations, such as reassociation and reciprocal approximations, and thereby improve numerical reproducibility.

All evolved fields are stored as multidimensional \kokkos \texttt{View} containers that reside in the device memory.
The initial data are read in and copied to the device at startup.
The entire evolution is then carried out on the device through computational kernels dispatched by \kokkos.
The fields are copied back to the host only for diagnostic output and snapshots.

We organize the parallelism in two levels: (i) MPI distributes the domain across ranks, and (ii) \kokkos drives the threads within each rank.
Specifically, the domain is split into a Cartesian grid of ranks along the three spatial directions.
Every refinement level is decomposed by this same grid of ranks, so each rank holds the same relative portion of every level.
By default we assume a symmetry about the equatorial plane, and we place ranks only in the upper half space.
Each rank holds a subdomain padded by a ghost buffer of six cells on every side.
This buffer is exchanged with the neighboring ranks by packing the boundary layers into contiguous arrays and communicating them with non blocking MPI calls.
These buffers reside in device memory and are handed directly to GPU-aware MPI, eliminating any staging through host memory.
Moreover, the same mechanism transfers data between refinement levels for the prolongation, restriction, exchange, and refluxing.

On the AMD APU, we made some specific optimizations.
The stencil kernels exploit the local data share of the device through the \kokkos team scratch mechanism.
These kernels are the finite difference derivatives, the KO dissipation, and the shift advection.
Specifically, each team loads one $8^3$ tile of a field, together with its halo, into the fast scratch memory.
The threads of the team then reuse it for the stencil, which avoids repeated reads from the high bandwidth memory.
This path is numerically identical to the default one and changes only the memory access pattern.

\section{Validation}\label{sec:validation}

To validate the correctness of \sacrak, we present in this section a suite of tests: a consistency check against the \sacrampi Fortran code (Section~\ref{sec:consistency}), a $\pi$ symmetry preservation test (Section~\ref{sec:pisym}), and a self convergence test for the BNS merger simulation (Section~\ref{sec:convergence}).
All of the \sacrak runs in this section are performed on the Viper APU cluster of the Max Planck Computing and Data Facility (MPCDF) described in Section~\ref{sec:performance}.

The initial data used in this section are generated by \fuka \citep{Papenfort:2021hod}.
We quote the resolution as $N$: each MPI rank holds $N^3$ active cells, so that under the $2\times2\times1$ rank decomposition a refinement level contains $2N \times 2N \times N$ active cells.
For example, $N = 40$ corresponds to a level of $80 \times 80 \times 40$ active cells.
We fix the size of the finest domain first, and obtain the grid spacing from this size and the cell count.
The finest domain encloses each compact object with a margin: it spans $1.2$ times the neutron star diameter for the BNS and BHNS, and $2\,m_0$ for the BBH.
In the BHNS case the domain that covers the black hole has the same size as that of the neutron star.
For all runs we set the Z4c damping parameters to $\kappa_1 = 0.005\,M_\odot^{-1}$ and $\kappa_2 = 0$, and the shift damping parameter to $\eta = 0.36\,M_\odot^{-1}$, unless otherwise specified.

\subsection{Consistency with \sacrampi Fortran}\label{sec:consistency}

\sacrak is a direct port of the \sacrampi Fortran code.
Given identical inputs, it should ideally reproduce the output of the parent code to within floating point round off.
In practice, however, bitwise equivalence cannot be reached.
The two codes are compiled with different toolchains and math libraries and run on different hardware.
These differences alter the order of floating point operations, and thus the round off pattern.

We therefore deem the port successful if the residual difference between the two codes is below three references: the typical spread between independent NR codes evolving the same physical configuration, the error between different resolutions with the same code, and the distinguishability threshold of current ground-based gravitational-wave detectors \citep{Lindblom:2008cm}.
For the former, cross-code comparisons typically find accumulated phase differences of $\mathcal{O}(0.1)$--$\mathcal{O}(1)\,\mathrm{rad}$ and fractional amplitude differences of $\mathcal{O}(0.01)$--$\mathcal{O}(0.1)$ at merger, corresponding to noise-weighted mismatches of $10^{-4}$--$10^{-3}$ \citep{Hannam:2009hh,Hinder:2013oqa,Lovelace:2016uwp,Radice:2025djo,Kuan:2025bzu,Hamilton:2024ziw}.
Resolution studies have shown that the inter-resolution phase difference accumulated up to merger can be $\mathcal{O}(0.1)\,{\rm rad}$ \citep{Kiuchi:2017pte,Kiuchi:2019kzt,Scheel:2025jct}.
For the latter, two waveforms cannot be distinguished by a detector if their noise-weighted mismatch satisfies $\bar{\mathcal{M}} \lesssim \bar{D}/(2\tilde{\rho}^{2})$ \citep{Flanagan:1997kp,Lindblom:2008cm}, where $\tilde{\rho}$ is the matched-filter signal-to-noise ratio and $\bar{D} \geq 1$ accounts for the number of measured parameters \citep{Chatziioannou:2017tdw,Purrer:2019jcp}.
Since we assess the consistency of the two codes directly in terms of the relative amplitude difference $|\Delta A|/A$ and the GW phase difference $|\Delta\phi_{\rm GW}|$ of $h^{2,2}_+$, rather than through mismatches, we recast this criterion using $\bar{\mathcal{M}} \approx \left(\langle\Delta\phi_{\rm GW}^{2}\rangle+\langle(\Delta A/A)^{2}\rangle\right)/2$, valid for small deviations \citep{Lindblom:2008cm}, where $\langle\,\cdot\,\rangle$ denotes a noise-weighted average over the signal.
Indistinguishability is then guaranteed if both $|\Delta\phi_{\rm GW}|$ and $|\Delta A|/A$ remain below $\sim 1/\tilde{\rho}$.
Even for the loudest event observed to date, GW250114 with $\rho \approx 80$ \citep{LIGOScientific:2025rid}, this conservative criterion ($\bar{D}=1$) requires only $|\Delta\phi| \lesssim 10^{-2}\,\mathrm{rad}$ and $|\Delta A|/A \lesssim 10^{-2}$; for the loudest BNS signal, GW170817 with $\rho \approx 32$ \citep{LIGOScientific:2017vwq}, the corresponding bound is $\approx 3\times10^{-2}$.
Accordingly, we require the residuals between \sacrak{} and \sacrampi{} to satisfy $|\Delta\phi_{\rm GW}| \leq 10^{-2}\,\mathrm{rad}$ and $|\Delta A|/A \leq 10^{-3}$ throughout the inspiral up to the merger (dotted lines in Figs.~\ref{fig:consistency_bbh}--\ref{fig:consistency_bns}), an order of magnitude below the typical inter-code/inter-resolution spread and at or below the distinguishability threshold of the loudest signal observed to date.

Specifically, we verify this for three representative binary configurations: BBH, BHNS, and BNS.
On both codes we use exactly the same initial data, grid hierarchy, and MPI domain decomposition.
We deliberately adopt a relatively low resolution of $N=40$ to make the test stringent.
Any subtle algorithmic or floating point inconsistency between the two implementations then accumulates rapidly and becomes visible in the waveform comparison.
The two evolutions are expected to remain in close numerical agreement throughout the inspiral.
We therefore apply no time or phase alignment, and all differences shown below are obtained by direct subtraction of the two waveforms.

\subsubsection{BBH}

\begin{figure}[tbp]
\centering
\includegraphics[width=\columnwidth]{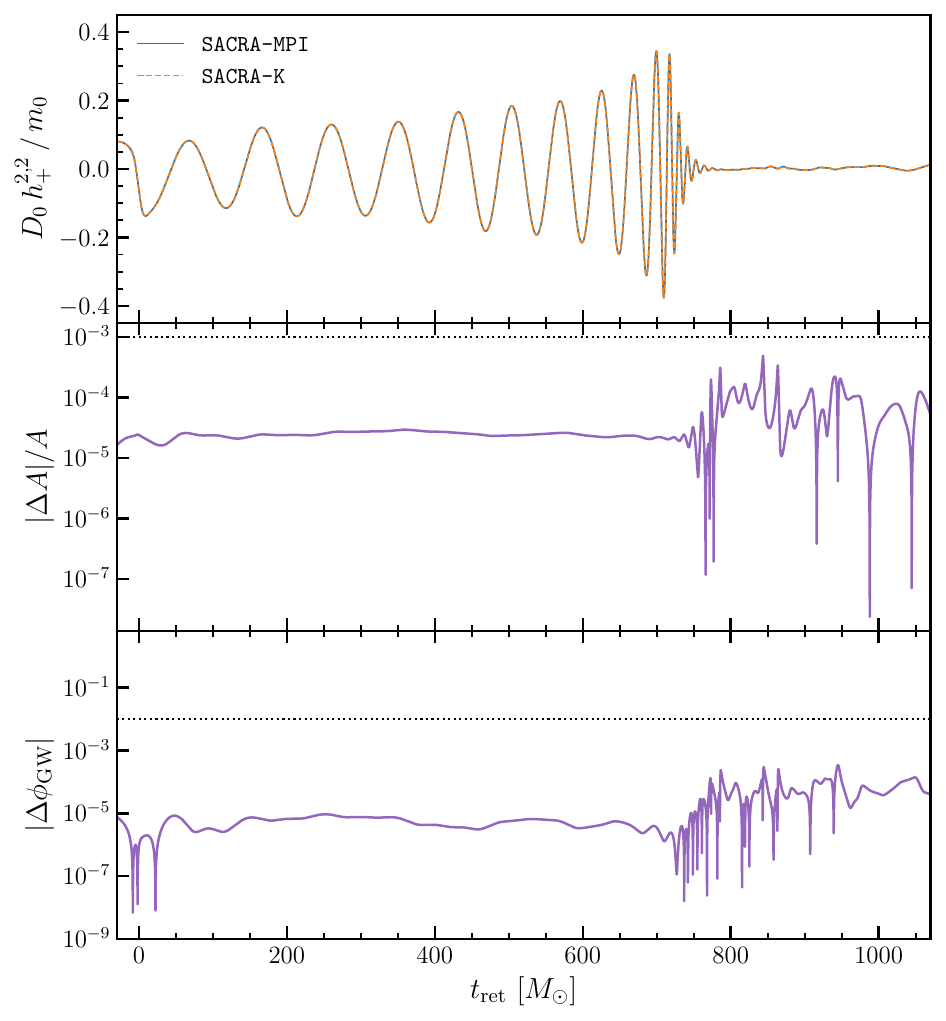}
\caption{Waveform comparison between \sacrak and \sacrampi Fortran for the BBH case.
\textit{Top:} real part of the $l = m = 2$ mode of $h^{l,m}$ extracted on a sphere, for \sacrampi Fortran (blue solid) and \sacrak (orange dashed).
\textit{Middle:} relative amplitude difference $|\Delta A|/A$.
\textit{Bottom:} absolute GW phase difference $|\Delta\phi_{\rm GW}|$.
The horizontal black dotted lines denote the acceptance criteria, $10^{-3}$ for the relative amplitude difference and $10^{-2}\,{\rm rad}$ for the phase difference.}
\label{fig:consistency_bbh}
\end{figure}

For the BBH consistency test we adopt an equal mass non spinning BBH with an initial separation of $d_0 = 10\,m_0$ and total mass is supposed to be $m_0 = 1\,M_\odot$.
The real part of the $(2,2)$ mode of the GW strain $h^{2,2}_+$ is extracted on a sphere of radius $r_0 = 24\,M_\odot$.
Figure~\ref{fig:consistency_bbh} compares the waveforms produced by the two codes, where we can see that the waveform is composed of the inspiral, merger, and ringdown~\citep{Buonanno:2006ui}.
Note that this extraction radius is not sufficiently large for a scientific run, since it is only $2.4$ times the initial separation of the binary. 
We nevertheless adopt this radius in the present test because, in this low-$N$ setup, resolving the high frequency ringdown signal at the extraction sphere requires a sufficiently fine grid spacing. 
Reducing the extraction radius is therefore the only practical way to maintain adequate resolution in the extraction region. 
Since the same extraction radius and the same analysis procedure are used for both codes, the dominant finite radius bias is expected to be largely common mode and may partially cancel in the code to code comparison. 
The measured discrepancy may therefore be interpreted as a possible underestimate of the total error with respect to the asymptotic waveform at null infinity

As shown in Figure~\ref{fig:consistency_bbh}, the strains are visually indistinguishable across the entire inspiral and merger.
Specifically, the relative amplitude difference $|\Delta A|/A$ stays below $\sim 10^{-4}$ throughout the inspiral phase, and increase a little but still below $\sim10^{-3}$ during merger and ringdown.
The GW phase difference $|\Delta\phi|$ is below $\sim 10^{-5}\,{\rm rad}$ in the inspiral phase, and increases to $\sim10^{-3}\,{\rm rad}$ in merger and ringdown.
These residuals are several orders of magnitude smaller than the typical spread reported in published code comparison studies \citep{Ferguson:2023vta,Radice:2025djo} or resolution studies \citep{Rashti:2024yoc,Scheel:2025jct} of BBH inspirals.
We therefore conclude that \sacrak passes the BBH consistency test.

\subsubsection{BHNS}

\begin{figure}[tbp]
\centering
\includegraphics[width=\columnwidth]{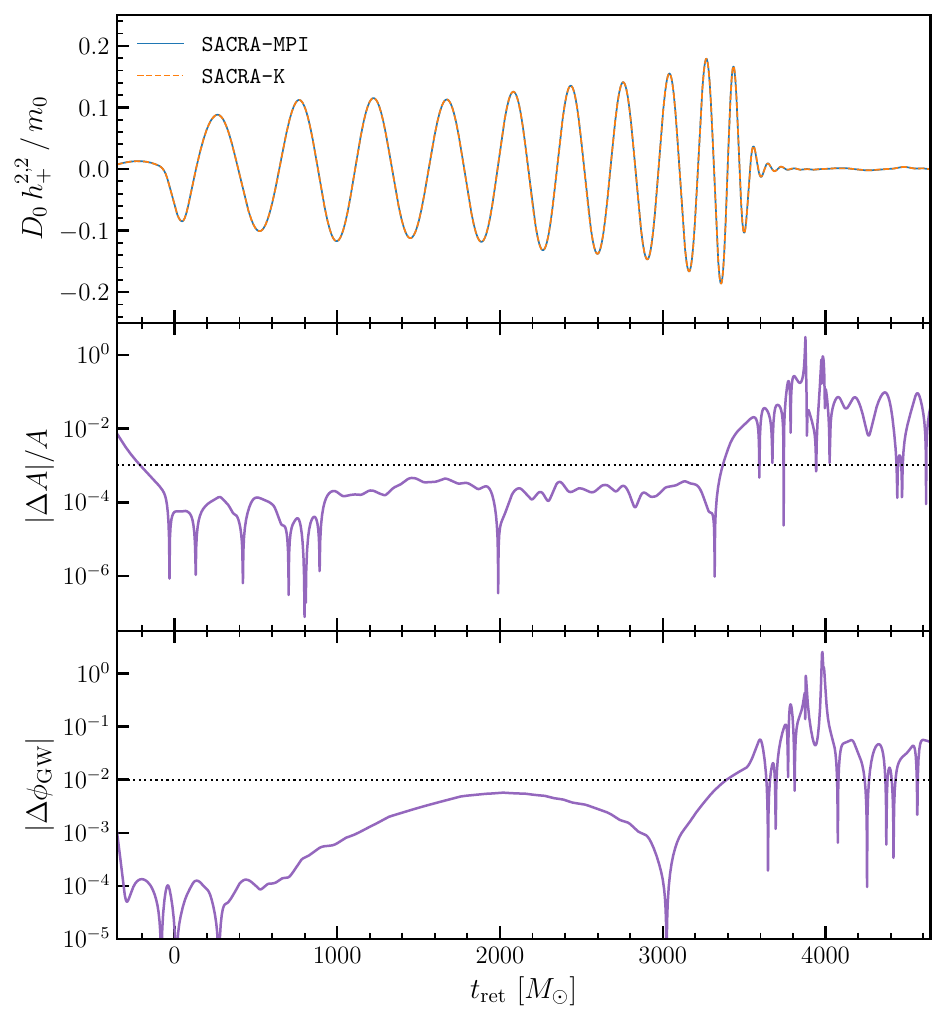}
\caption{Same as Figure~\ref{fig:consistency_bbh}, but for a BHNS
configuration.}
\label{fig:consistency_bhns}
\end{figure}

For the BHNS test the mass of the neutron star is $1.35\,M_\odot$, and the mass ratio is $5$.
The neutron star matter is described by the DD2 EOS \citep{Typel:2009sy,Hempel:2009mc} in tabulated form, with $\Gamma_{\rm th} = 1.8$.
The Christodoulou mass of the black hole is $6.75\,M_\odot$, and the dimensionless spin is $0.75$, aligned with the orbital angular momentum.
The binary starts from an initial separation $d_0 = 55\,M_\odot$, and GW strain is extracted on a sphere of radius $r_0 = 300\,M_\odot$.
For this model, the neutron star is tidally disrupted at an orbit close to the innermost stable circular orbit. Thus, a disk with small mass is formed around the remnant black hole although a substantial fraction of the neutron star matter falls simultaneously into the black hole, exciting a quasi-normal mode of the black hole.

Figure~\ref{fig:consistency_bhns} compares the two waveforms.
As in the BBH case, the real parts of $h^{2,2}$ are visually indistinguishable across the entire simulation.
Specifically, the relative amplitude difference $|\Delta A|/A$ is ($\lesssim 10^{-3}$) through the inspiral ($t_{\rm ret} \lesssim 3300\,M_\odot$), and grows to $\gtrsim 10^{-1}$ only after merger.
The GW phase difference $|\Delta\phi_{\rm GW}|$ stays $\lesssim 10^{-2}\,{\rm rad}$ throughout the inspiral and likewise grows to $\gtrsim 10^{-1}\,{\rm rad}$ after merger.
These residuals sit at the acceptance criterion: $|\Delta A|/A$ rides on the $10^{-3}$ line through the late inspiral, and $|\Delta\phi_{\rm GW}|$ touches $10^{-2}\,{\rm rad}$ just before merger, so the BHNS case meets the requirement of this section only marginally.
They are larger than in the BBH case, as expected.
The hydrodynamic part introduces sharp flow features, in particular the stellar surface.
The finite volume reconstruction of such features is far more sensitive to the order of floating point operations than the smooth spacetime evolution, and any such difference is amplified through the strongly nonlinear dynamical process.
During the inspiral, however, the residuals remain below the distinguishability threshold of the current GW detector.
Moreover, after the final stage of the ringdown, the amplitude is so low that the effects of truncation errors associated with floating point operations can be significant in this problem.
We therefore conclude that \sacrak passes the BHNS consistency test. 

\subsubsection{BNS}\label{sec:consis_bns}

\begin{figure}[tbp]
\centering
\includegraphics[width=\columnwidth]{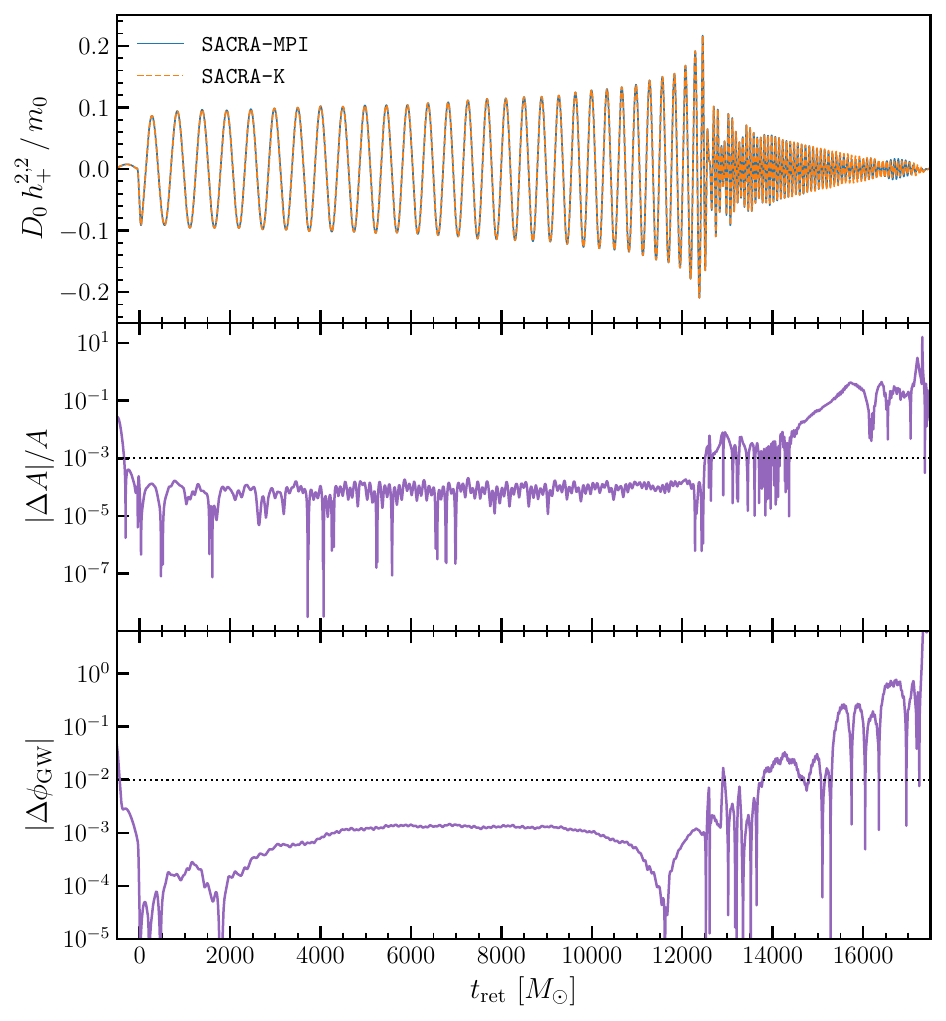}
\caption{Same as Figure~\ref{fig:consistency_bbh}, but for a BNS
configuration.}
\label{fig:consistency_bns}
\end{figure}

The BNS test is the most stringent of the three.
The inspiral is the longest, and the binary leaves behind a long lived remnant rather than promptly ringing down.
We evolve an equal-mass BNS with total mass $2.7\,M_\odot$ and initial separation $d_0 = 42\,M_\odot$.
The GW strain $h^{2,2}_+$ is extracted on a sphere of radius $r_0 = 480\,M_\odot$.
We use the DD2 EOS \citep{Typel:2009sy,Hempel:2009mc} and $\Gamma_{\rm th} = 1.8$.
The merger occurs near $t_{\rm ret} \approx 12000\,M_\odot$, preceded by approximately $15$ orbits.
For this model, a long-lived massive neutron star is formed as a remnant.
The slowly damped oscillations of the remnant persist to the end of the simulation at $t_{\rm ret} \approx 18000\,M_\odot$ at which the remnant neutron star still survives.

Figure~\ref{fig:consistency_bns} compares the two waveforms.
The real parts of $h^{2,2}$ track each other across the entire inspiral and through merger.
A divergence becomes visible by eye only in the late post-merger signal ($t_{\rm ret} \gtrsim 16000\,M_\odot$).
Specifically, the relative amplitude difference $|\Delta A|/A$ sits in the $10^{-5}$--$10^{-3}$ band throughout the long inspiral and grows to $\gtrsim10^{-1}$ during the post-merger phase.
The GW phase difference $|\Delta\phi_{\rm GW}|$ remains at the $\sim 10^{-3}\,{\rm rad}$ level across the inspiral, then grows to $10^{-1}\,{\rm rad}$ in the early post-merger phase, and finally increase to $O(1)\,{\rm rad}$.
This post-merger growth is expected.
The differentially rotating remnant is a highly nonlinear, turbulent system, in which round off level differences between the two implementations are amplified exponentially.
The inspiral and merger primarily carry the waveform information relevant for current and future GW detectors.
There, the residuals stay far below the spread between independent NR codes \citep{Hamilton:2024ziw,Habib:2025bkb} and different resolutions of the same code \citep{Kiuchi:2017pte,Kiuchi:2019kzt}, so we conclude that \sacrak passes the BNS consistency test.

As shown in this section, the two codes do not agree quantitatively in the long-term evolution of the merger remnant.
At merger, the collision of the two stars launches strong shocks, across which the reconstruction locally degrades to first order.
In addition, the shear layer at the contact interface is susceptible to the Kelvin–Helmholtz instability \citep{Kiuchi:2015sga,Kiuchi:2026pgb}, which can amplify small differences and contribute to the rapid growth of post-merger phase differences.
The remnant therefore enters a strongly nonlinear, effectively chaotic regime.
Perturbations at the round off level, whose spatial pattern can depend on the hardware, compiler, floating point contraction, reduction order, and numerical libraries, may be amplified by shocks, shear instabilities, and turbulent mixing.
Consequently, two executables that are not bitwise identical can decorrelate during the post-merger evolution, even if they agree during the inspiral and early merger.
This behavior suggests that numerically obtaining high precision waveforms for the post-merger phase of BNS mergers may be intrinsically challenging.

\subsection{\texorpdfstring{$\pi$}{Pi} symmetry preservation}\label{sec:pisym}

\begin{figure}[tbp]
\centering
\includegraphics[width=\columnwidth]{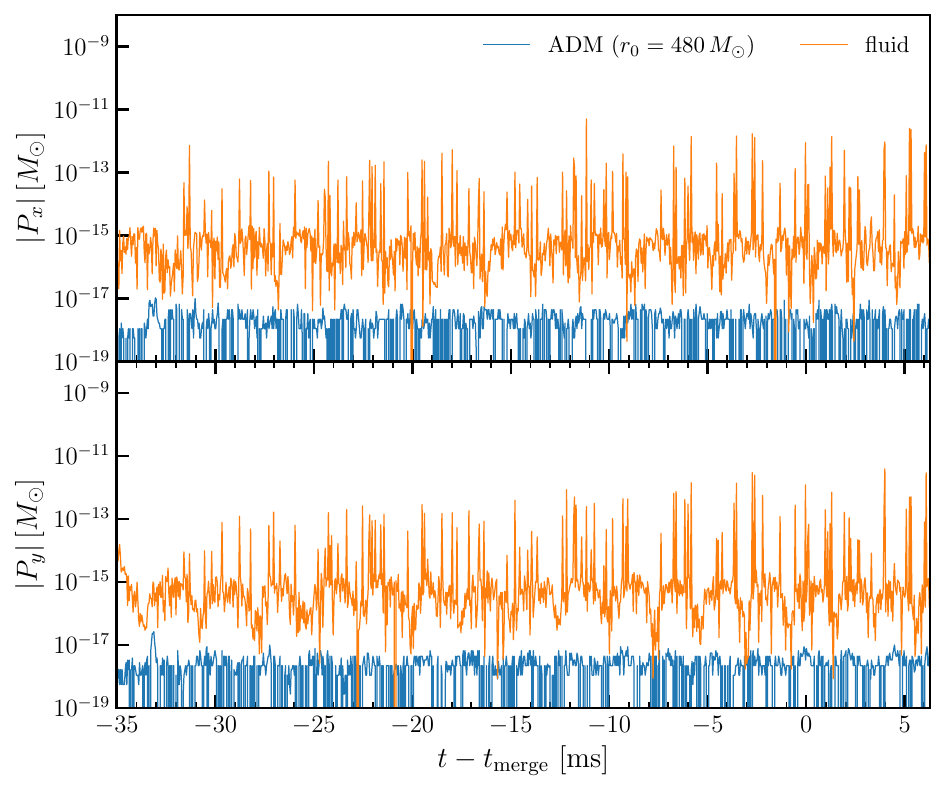}
\caption{$\pi$ symmetry preservation diagnostic for an equal mass non spinning BNS evolution at resolution $N = 102$.
The momentum components $|P_x|$ and $|P_y|$ are shown in the upper and lower panels; in each panel the ADM and fluid linear momenta are drawn in blue and orange, respectively.}
\label{fig:pi_symmetry}
\end{figure}

We follow the $\pi$ symmetry preservation test of \citet{Kiuchi:2025ksk,Gao:2025nfj}, where the detailed definition of the symmetry can be found.
We use the same initial data as the BNS consistency test of Sec.~\ref{sec:consis_bns}, but at resolution $N = 102$.
The initial data generated by \fuka are not guaranteed to be bitwise $\pi$ symmetric.
We therefore symmetrize them before evolving: the field value at each point $(x, y, z)$ in a half of the whole domain is replaced by the value at its symmetry partner $(-x, -y, z)$, multiplied by the sign appropriate to the field component for vectors and tensors.

For this test we additionally compile with the \texttt{-ffp-contract=off} flag to ensure that the floating point operations are carried out in the intended order.
By default the compiler may contract a multiplication followed by an addition into a single fused multiply add (FMA) with a single rounding.
When this contraction is applied differently to the two members of a symmetric pair, it breaks their bitwise correspondence, and hence the $\pi$ symmetry, at the round off level.

During the evolution we monitor the symmetry directly.
For every symmetric pair of points, we compute the difference between each field and its symmetry mapped partner, and we record the maximum absolute difference over all grid points and variables.
This maximum remains exactly zero, to the last bit, throughout the entire simulation.
This confirms that neither the algorithm nor its \kokkos parallelization breaks the symmetry.

As an integrated diagnostic we also monitor the in-plane linear momentum, computed in two independent ways.
Both vanish identically for a $\pi$ symmetric configuration.
Specifically, the fluid linear momentum is the volume integral
\begin{equation}
P^{\rm fluid}_{x/y} = \int \sqrt{\gamma}\, D\, h\, u_{x/y}\, d^3x,
\label{eq:P_fluid}
\end{equation}
where $\gamma$ and $u_{x/y}$ are the determinant of the three spatial metric and the $x/y$ component of the four velocity, respectively, and the ADM linear momentum is the surface integral
\begin{equation}
P^{\rm ADM}_i(r) = \frac{1}{8\pi}\oint_{S_r}\,
\left(K_i{}^j - \gamma_i{}^j K\right)dS_j,
\label{eq:P_ADM}
\end{equation}
evaluated on the coordinate sphere of radius $r = r_0 = 480\,M_\odot$.

As shown in Figure~\ref{fig:pi_symmetry}, both diagnostics oscillate at the double precision round off floor throughout the evolution, $\sim 10^{-15}$ for the fluid linear momentum and $\sim 10^{-18}$ for the ADM linear momentum.
These oscillations arise because the order of the \kokkos parallel reduction (here a summation) is not fixed.
The exactly cancelling contributions are therefore accumulated in a different floating point order at each output.
To confirm that they do not mask a slow symmetry breaking, we evolve the configuration for a long time.
The residual remains stationary at this small level and shows no growth with time.
We therefore conclude that \sacrak passes the $\pi$ symmetry test, and that the port introduces no errors capable of breaking the symmetry.

\subsection{Self convergence test of the BNS merger}\label{sec:convergence}

The previous sections established that \sacrak reproduces \sacrampi to high accuracy and introduces no errors that break the $\pi$ symmetry.
We now present a self convergence test that quantifies how the results converge with increasing resolution.
We perform this test only for the BNS case, since it exercises both the spacetime and the hydrodynamic evolution.
We use the same initial data as \citet{Kiuchi:2025ksk} to allow a direct comparison of the convergence behaviour; the detailed settings can be found there.
The configuration is an equal mass, non spinning binary with $m_1 = m_2 = 1.35\,M_\odot$, the H4 EOS \citep{Glendenning:1991es} in a four segment piecewise polytropic representation with $\Gamma_{\rm th} = 5/3$, and an initial orbital angular velocity of $m_0\Omega_0 \approx 0.0161$, which corresponds to about $12$--$13$ orbits before merger.
The shift damping parameter is set to be $\eta = 0.03\,M_\odot^{-1} \approx 0.1/m_0$, in order to keep consistent with \citet{Kiuchi:2025ksk}.
The four resolutions are $N = 92, 108, 124, 188$, which correspond to $N = 93, 109, 125, 189$ in \citet{Kiuchi:2025ksk}.
The offset by one reflects the different cell counting conventions of the two codes; the grids and their spacings are identical.

To quantify the convergence we adopt the standard ansatz for the GW phase,
\begin{equation}
\phi_{\rm GW}(t; N) = \phi_{\rm GW}^{\infty}(t)
+ a(t)\left(\frac{\Delta x_N}{\Delta x_{188}}\right)^{p_{\rm conv}(t)},
\label{eq:conv_ansatz}
\end{equation}
where $\Delta x_N$ is the finest grid spacing at resolution $N$, so that $\Delta x_{188}$ is that of the highest resolution, $\phi_{\rm GW}^{\infty}$ is the continuum limit phase, $a$ is the amplitude of the leading error term, and $p_{\rm conv}(t)$ is the local convergence order.
We determine the three unknowns $\phi_{\rm GW}^{\infty}$, $a$, and $p_{\rm conv}$ by a least squares fit to the measured phases $\phi_{\rm GW}(t; N)$ at all four resolutions.

\begin{figure}[tbp]
\centering
\includegraphics[width=0.45\textwidth]{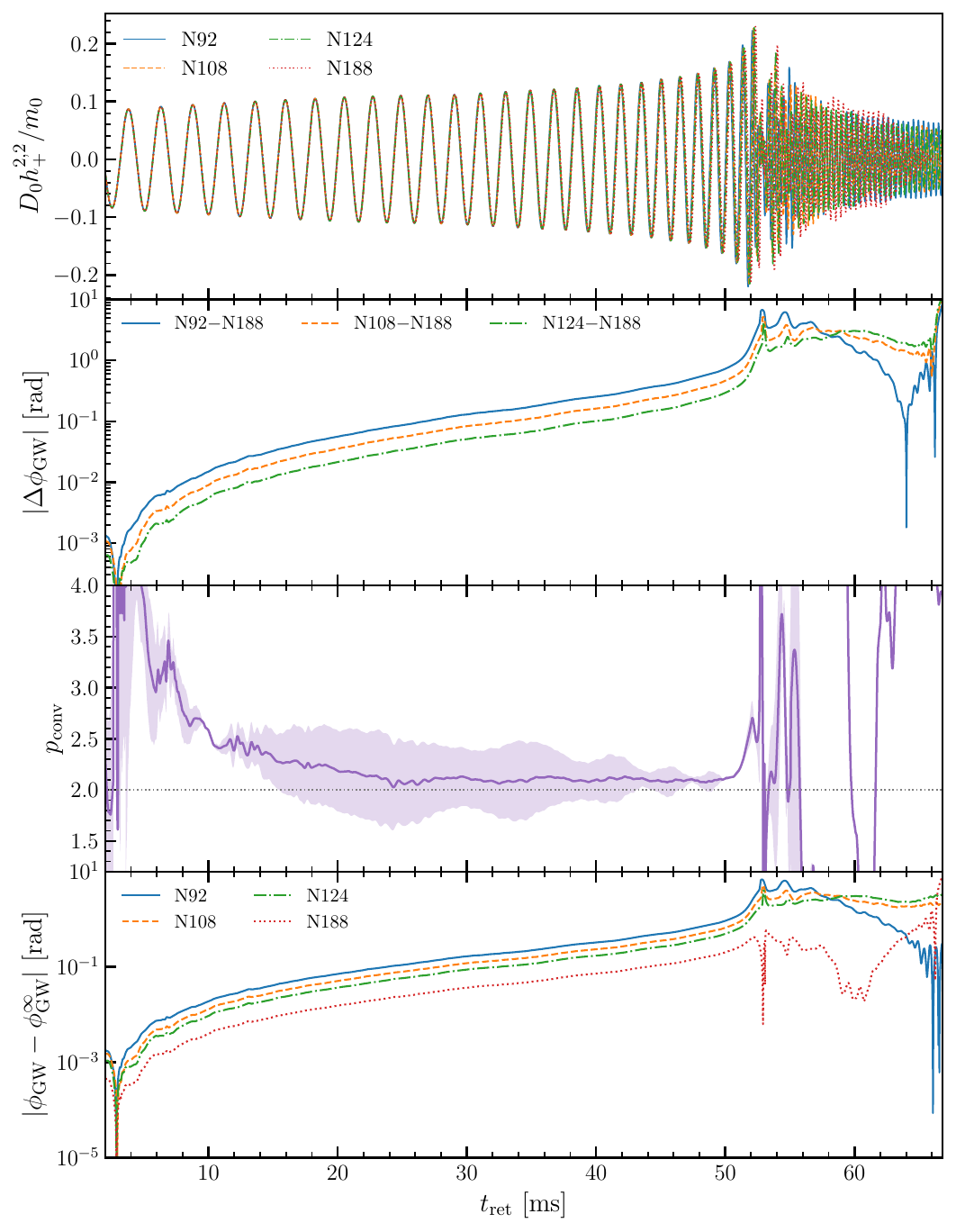}
\caption{Self convergence of the \sacrak BNS merger simulation at four resolutions, $N = 92$ (blue solid), $108$ (orange dashed), $124$ (green dash dotted), and $188$ (red dotted), shown against the
retarded time $t_{\rm ret}$.
\textit{Top:} dominant strain mode $D_0h_+^{22}/m_0$.
\textit{Second:} gravitational wave phase difference $|\Delta\phi_{\rm GW}|$ of the three lower resolutions relative to $N = 188$.
\textit{Third:} locally fitted convergence order $p_{\rm conv}$, with the shaded band showing the fit uncertainty and the dotted line marking $p_{\rm conv} = 2$.
\textit{Bottom:} absolute difference between the GW phase $\phi_{\rm GW}$ at each resolution and the extrapolated phase $\phi_{\rm GW}^{\infty}$.}
\label{fig:convergence}
\end{figure}

The results are shown in Figure~\ref{fig:convergence}.
The top panel shows the dominant strain mode $D_0h_+^{2,2}/m_0$ at the four resolutions.
The waveforms overlap throughout the inspiral and begin to separate only near the merger at $t_{\rm ret} \approx 50\,\mathrm{ms}$.
The second panel shows the GW phase difference $|\Delta\phi_{\rm GW}|$ of the three lower resolutions relative to the highest, $N = 188$.
The differences decrease monotonically with increasing resolution; in time, they grow from $\sim 10^{-3}\,\mathrm{rad}$ in the early inspiral to $\sim 10^{-1}\,\mathrm{rad}$ shortly before merger and reach order unity at merger.
The third panel shows the locally fitted convergence order $p_{\rm conv}$ from Equation~\eqref{eq:conv_ansatz}.
It stabilizes around $p_{\rm conv} \approx 2$ throughout the inspiral.
The metric uses fourth order finite differencing and the fluid a third order PPM reconstruction.
The global convergence of the hydrodynamics, however, is limited to second order by the approximate Riemann solver and the stellar surface.
Near merger, $p_{\rm conv}$ becomes noisy as the waveform develops sharp features.
The bottom panel shows the difference between the GW phase $\phi_{\rm GW}$ at each resolution and the extrapolated phase $\phi_{\rm GW}^{\infty}$.
This difference likewise shrinks with increasing resolution.
Together, these panels demonstrate that \sacrak converges at the expected second order rate throughout the BNS inspiral.
This matches the convergence order of $\approx 2$ that \citet{Kiuchi:2025ksk} reports for the same configuration evolved with \sacrampi and the same PPM based Riemann solver.

\section{Performance}\label{sec:performance}

We developed \sacrak to bring GPU acceleration to \sacrampi.
The central question is therefore how much faster it runs than the original Fortran code on a CPU.
We answer this question in two steps: (i) we first compare \sacrak on accelerators against \sacrampi on a CPU in the smallest configuration, and (ii) we then measure how \sacrak scales across many devices.

All performance tests were carried out on the MPCDF Raven and Viper clusters.\footnote{\url{https://docs.mpcdf.mpg.de/doc}}
The Fortran \sacrampi runs on the Raven CPU cluster, where each node holds two Intel Xeon IceLake Platinum 8360Y CPUs with 72 cores in total.
\sacrak, in turn, runs on two accelerator platforms: (i) the Raven GPU cluster, with four NVIDIA A100 GPUs and two of the same CPUs per node, and (ii) the Viper APU cluster, with two AMD Instinct MI300A APUs and 48 CPU cores per node.

\subsection{Smallest configuration performance and architecture comparison}\label{sec:perf_single_node}

Large simulations distributed across many nodes pay an MPI communication cost.
This cost lowers the scaling efficiency and obscures the raw speed of the code.
To measure how much faster the \sacrak port runs than the original Fortran \sacrampi, we begin with the smallest configuration that still exercises the full algorithm.
This keeps the communication overhead as small as possible.
Specifically, this smallest size test uses an MPI decomposition of $2\times2\times1$, i.e., two ranks along each of the $x$ and $y$ directions and one along $z$, for a total of four ranks.
We run the Fortran \sacrampi on Raven CPU and \sacrak on the two accelerators, mapping the four ranks onto each platform in the way that best matches its hardware.
Specifically, Raven CPU uses a single node with 18 cores per rank, Raven GPU a single node with one A100 device per rank, and Viper APU two nodes with one MI300A per rank.
The timed problem is a DD2 BNS configuration with the same settings as a production run, including all diagnostic output.
For each platform and resolution we report the zone cycles per second (ZCPS), normalized per accelerator device or per CPU core, averaged over 20 iterations, in Table~\ref{tab:single_node_platforms}.
Here one iteration denotes one coarsest level step; its zone cycle count is the number of level evolutions performed within it multiplied by the active cells per level, and ZCPS is this count divided by the wall clock time of the iteration.

\begin{table}[tbp]
\centering
\footnotesize
\caption{ZCPS per GPU/APU device / CPU core for the smallest configuration performance comparison, with the Fortran \sacrampi on the CPU and \sacrak on the two accelerators.
}
\label{tab:single_node_platforms}
\begin{tabular}{lcccc}
\hline\hline
Platform & $N=60$ & $N=80$ & $N=100$ \\
\hline
Single Intel 8360Y core & $2.14\times 10^4$ & $2.24\times 10^4$ & $2.64\times 10^4$ \\
NVIDIA A100 & $4.20\times 10^6$ & $5.89\times 10^6$ & $6.56\times 10^6$ \\
AMD MI300A & $3.60\times 10^6$ & $5.76\times 10^6$ & $7.30\times 10^6$ \\
\hline
\end{tabular}
\end{table}

We first compare the throughput per core or per device.
At $N=100$, for instance, a single CPU core sustains $2.64\times 10^4$ ZCPS, while a single A100 sustains $6.56\times 10^6$ ZCPS and a single MI300A $7.30\times 10^6$ ZCPS.
One accelerator device therefore delivers the throughput of roughly $250$ to $280$ CPU cores.
In practice, however, simulations rarely run on a single core.
A fairer comparison is between the smallest configurations as a whole: one node of the Raven CPU cluster with 72 cores, one node of the Raven GPU cluster with four A100s, and two nodes of the Viper APU cluster with four MI300As.
At $N=100$, these sustain $1.90\times10^6$, $2.62\times10^7$, and $2.92\times10^7$ ZCPS, respectively.
The speedups are then $13.8\times$ on the A100 and $15.4\times$ on the MI300A.
Moreover, this advantage widens with resolution, from about $10\times$ at $N=60$ to $14$--$15\times$ at the larger grids.
The accelerators are underused at small problem sizes and need a larger grid to be saturated.
On either accelerator, \sacrak thus outperforms the Fortran \sacrampi on the CPU by roughly an order of magnitude.

In practice, the two accelerators are closely matched, despite their different specifications on paper.
Specifically, the A100 is slightly faster at the smaller resolutions ($4.20\times 10^6$ against $3.60\times 10^6$ ZCPS at $N=60$ and $5.89\times 10^6$ against $5.76\times 10^6$ ZCPS at $N=80$).
The MI300A, however, pulls ahead at $N=100$ ($7.30\times 10^6$ against $6.56\times 10^6$ ZCPS).
The MI300A holds a large theoretical advantage.
Its peak FP64 vector throughput of $61.3\,$TFLOPS is about six times the $9.7\,$TFLOPS of the A100.
However, this advantage does not carry over to the measured throughput.
We attribute the gap to differences between the CUDA and HIP compilers and between the cache hierarchies of the two architectures.
Moreover, the unified 128 GB memory of the MI300A lets it reach $N=144$, where it sustains $7.72\times 10^6$ ZCPS per device.

\subsection{Strong scaling}\label{sec:perf_strong_scaling}

\begin{figure}[tbp]
\centering
\includegraphics[width=\columnwidth]{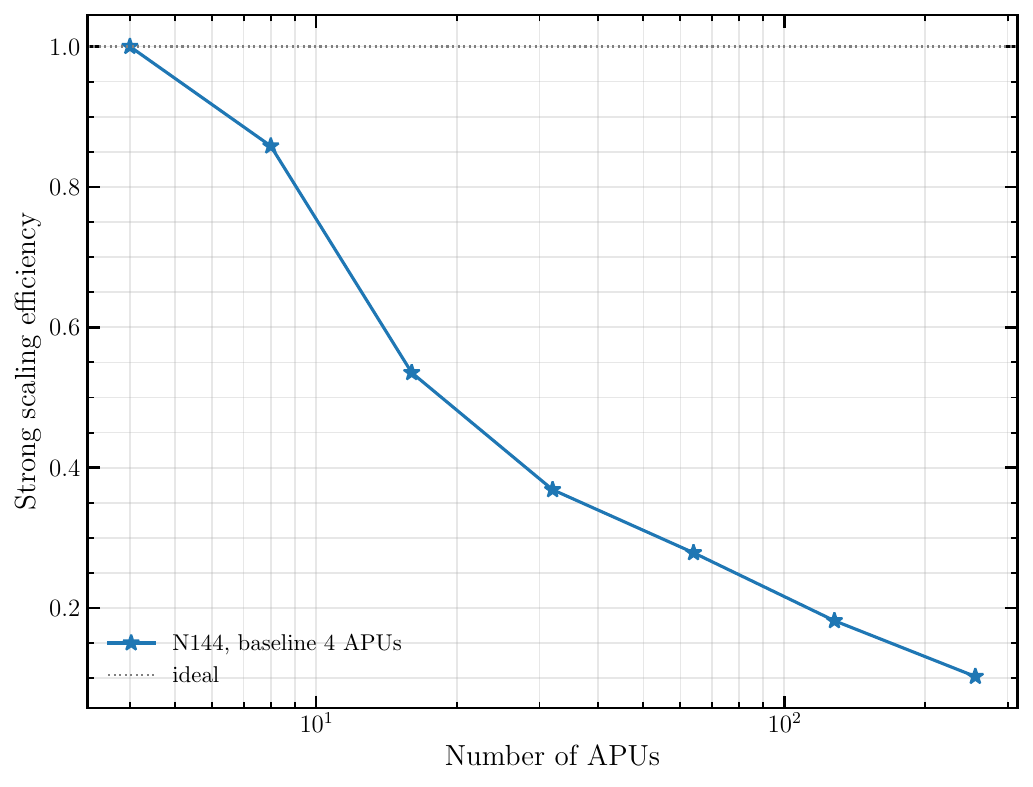}
\caption{Strong scaling of \sacrak on the Viper APU cluster at a fixed total problem size ($N=144$).
Stars show the strong scaling efficiency, the achieved speedup divided by the ideal linear speedup, against the number of APU devices, from the four device $2\times2\times1$ baseline to 256 devices.
The dotted line marks ideal scaling.}
\label{fig:strong_scaling}
\end{figure}

We next measure the strong scaling of \sacrak on the Viper APU cluster.
We fix the total problem at the $N=144$ configuration and distribute it over a growing number of devices.
Specifically, we take a $2\times2\times1$ decomposition on four APU devices across two nodes as the baseline, and double the device count in steps up to 256 devices on 128 nodes.
Figure~\ref{fig:strong_scaling} then reports the strong scaling efficiency, the achieved speedup divided by the ideal linear speedup, referenced to this four device baseline.

The throughput rises with device count but increasingly falls short of the ideal.
Specifically, relative to the four device baseline, the time per iteration drops from $13.92\,$s to $2.12\,$s at 256 devices.
This is a speedup of only $6.6\times$ for a $64\times$ increase in devices.
The parallel efficiency therefore falls steadily, from $86\%$ at eight devices to $54\%$ at 16, $28\%$ at 64, and about $10\%$ at 256.

This loss of efficiency is driven mainly by the growing ratio of buffer to active cells as each subdomain shrinks.
This occurs because every subdomain is padded on each side by a buffer of six cells.
A rank therefore stores $(N+12)^3$ cells to update $N^3$ active ones.
Specifically, at the four device baseline each rank holds $(144+12)^3$ cells to update $144^3$ active ones, so about $21\%$ are buffer.
By the $4\times4\times2$ decomposition at 32 devices, the subdomain has shrunk to $72^3$ active cells within $(72+12)^3$, which raises the buffer share to about $37\%$.
This redundant fraction keeps growing with the device count.
The communication of the buffer zones therefore comes to dominate the runtime.

\subsection{Weak scaling}\label{sec:perf_weak_scaling}

\begin{figure}[tbp]
\centering
\includegraphics[width=\columnwidth]{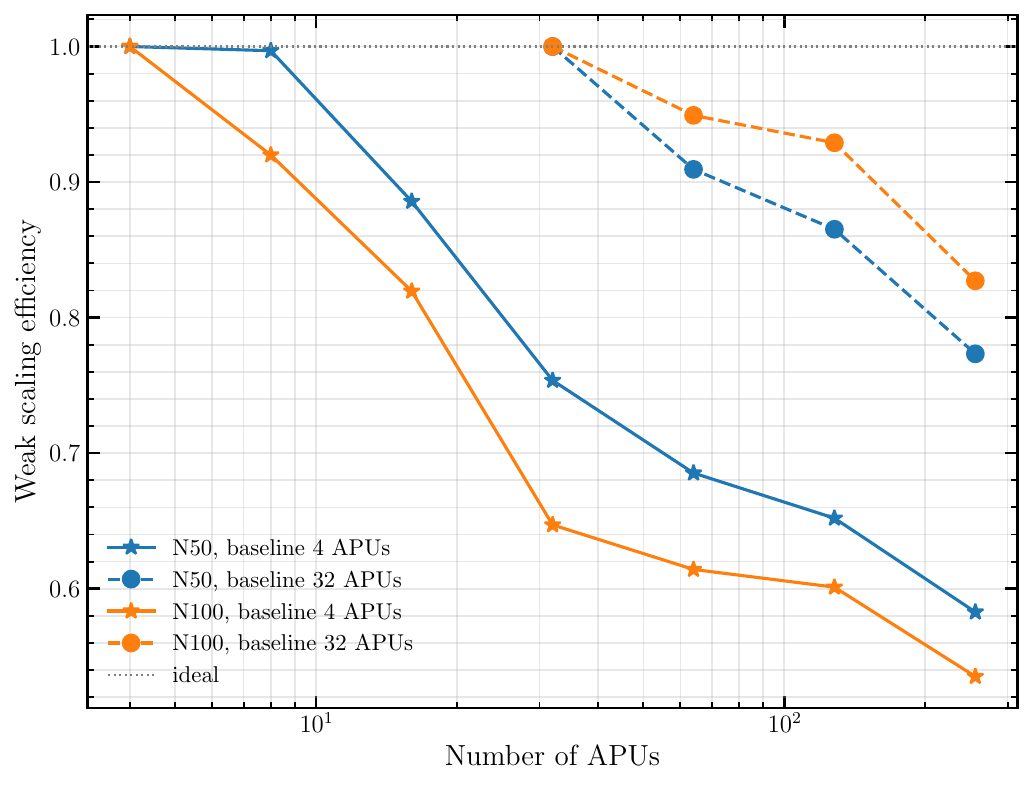}
\caption{Weak scaling of \sacrak on the Viper APU cluster at a fixed subdomain size per device.
The weak scaling efficiency $t_{\rm base}/t_n$ is shown against the number of APU devices for two per device loads, $N=50$ (blue) and $N=100$ (orange), each normalized to the four device baseline (solid) and to the 32 device baseline (dashed).
The dotted line marks ideal weak scaling.}
\label{fig:weak_scaling}
\end{figure}

We finally test the weak scaling of \sacrak on the Viper APU cluster.
We hold the subdomain on each device fixed, so that the total problem grows in proportion to the device count.
Specifically, we use two per device loads, $N=50$ and $N=100$ active cells along each direction, and scale from four to 256 devices.
The test again evolves the full AMR hierarchy with the production settings of Sec.~\ref{sec:perf_single_node}.
Under ideal weak scaling the time per iteration would stay constant.
We therefore quote the efficiency as the ratio $t_{\rm base}/t_n$ of the baseline time to the time at $n$ devices, shown in Figure~\ref{fig:weak_scaling}.

The obvious baseline, the four device $2\times2\times1$ run, is an unfair reference, because that decomposition has a single rank along $z$.
It therefore exchanges buffers in only two directions, and so runs faster per iteration than any configuration that communicates along all three axes.
Specifically, measured against it (solid curves), the efficiency falls to $58\%$ for $N=50$ and $54\%$ for $N=100$ at 256 devices.
We therefore also adopt the 32 device $4\times4\times2$ run as the representative baseline (dashed curves).
This run already communicates in all three directions.

Across the full 4-256 device range, the total problem size grows by a factor of 64.
Relative to the 32 device baseline, the largest run represents an 8 fold increase in both device count and total problem size.
Specifically, the efficiency stays above $90\%$ out to 64 devices and is still $77\%$ for $N=50$ and $83\%$ for $N=100$ at the largest run of 256 devices on 128 nodes.
Moreover, relative to this baseline, the larger per device load scales better at every device count.
More computation per device lowers the fraction of each step spent on communication.

\section{Summary}\label{sec:summary}

We have presented \sacrak, a C++ port of the Fortran code \sacrampi built with the \kokkos performance portability library.
Its main purpose is to bring GPU acceleration to the numerical relativity simulation of compact binaries.
At the same time, we maintain a single source code that runs on CPUs, GPUs, and APUs.

We validated the port against the parent code and found good agreement.
For the BBH, BHNS, and BNS configurations, the waveforms of \sacrak agree with those of \sacrampi to a level far below the spread between independent numerical relativity codes and at or below the distinguishability threshold of current detectors.
Moreover, the code preserves the $\pi$ symmetry to the last bit, and the BNS merger converges at the expected second order rate.

We then measured the performance.
In the smallest test configuration, \sacrak runs about an order of magnitude faster on either the A100 GPU or the MI300A APU than the Fortran \sacrampi does on the CPU.
We also measured its strong and weak scaling on the Viper APU cluster up to 256 devices.

Several directions remain for future work.
First, the code can be extended with more physics, in particular magnetohydrodynamics and neutrino radiation transport.
Second, overlapping the computation with the communication can further improve the parallel efficiency.
Finally, the GPU acceleration demonstrated here also opens the way to large surveys of the parameter space and to the construction of gravitational waveform libraries for compact binary mergers. 
Indeed, we have already performed a variety of simulations for BNS and binary strange stars with subsolar mass range to derive gravitational waveforms and to evaluate the dynamical ejecta mass. The result for this study is reported in a separate paper \citep{Gao:2026zwc}.

\begin{acknowledgments}

MH thanks Alan Tsz-Lok Lam, Carlo Musolino, Yong Gao, Hao-Jui Kuan, and Kota Hayashi for useful discussions.
Numerical computations were carried out on Sakura, Raven, and Viper at Max Planck Computing and Data Facility.
This work was in part supported by the National Natural Science Foundation of China under Grants No.~12233011, the Project for Young Scientists in Basic Research (Grant No.~YSBR-088) of the Chinese Academy of Sciences, and Grant-in-Aid for Scientific Research (Grants No.~23H04900 and No.~23K25869) of Japanese MEXT/JSPS.

\end{acknowledgments}

\bibliographystyle{aasjournalv7}
\bibliography{bibtex}{}
\end{document}